\documentclass[aps,preprint]{revtex4}%
\usepackage{amsfonts}
\usepackage{amsmath}
\usepackage{amssymb}
\usepackage{graphicx}%
\begin{document}
\title{Cold dilute neutron matter on the lattice II:\\Results in the unitary limit}
\author{Dean Lee and Thomas Sch{\"a}fer}
\affiliation{Department of Physics, North Carolina State University, Raleigh, NC 27695}

\begin{abstract}
This is the second of two papers which investigate cold dilute neutron matter
on the lattice using pionless effective field theory. \ In the unitary limit,
where the effective range is zero and scattering length is infinite, simple
scaling relations relate thermodynamic functions at different temperatures.
\ When the second virial coefficient is properly tuned, we find that the
lattice results obey these scaling relations. \ We compute the energy per
particle,\ pressure, spin susceptibility, dineutron correlation function, and
an upper bound for the superfluid critical temperature.

\end{abstract}
\maketitle


\section{Introduction}

\label{sec_int}

This is the second of two papers which investigate cold dilute neutron matter
on the lattice using pionless effective field theory. \ We refer to the first
paper as [I]. \ In this paper we study the scaling behavior of the results in
the unitary limit and compute a number of physical observables, the energy and
pressure, the spin susceptibility, and the dineutron correlation function.

In the unitary limit the scattering length is much larger than the
interparticle spacing whereas the range of the interaction is much smaller.
\ In this limit the only length scales in the problem are $(n_{f}(\mu
))^{-1/3}$ and the thermal wavelength $\lambda_{T}$. \ Here, $n_{f}(\mu
)=(3\pi^{2})^{-1}(2m_{N}\mu)^{3/2}$ is the density of a free Fermi gas and
$\lambda_{T}=\sqrt{2\pi/(m_{N}T)}$, where $m_{N}$ is the neutron mass, $\mu$
is the chemical potential, and $T$ is the temperature. \ All dimensionful
quantities can be expressed as suitable powers of either ($n_{f}(\mu))^{-1/3}$
or $\lambda_{T}$ times a function of the dimensionless quantity $\mu/T$. \ For
example, we can write the pressure as \cite{Ho:2004b}
\begin{equation}
P(T,\mu)=\frac{2}{5}\mu n_{f}(\mu)\mathcal{G}(x)
\end{equation}
where $x=\mu/T$. \ Using standard thermodynamic identities one can show that
\begin{equation}
\rho(\mu)=n_{f}(\mu)\left[  \mathcal{G}(x)-\frac{2x}{5}\mathcal{G}^{\prime
}(x)\right]  \text{, \ }\epsilon=\frac{3P}{2}\text{,}%
\end{equation}
where $\rho$ is the density and $\epsilon$ is the energy density. \ In this
work we first show how these relations arise in the lattice regularized
theory. \ We then present numerical results for the equation of state, the
spin susceptibility, and the dineutron correlation function, and show that the
lattice data are consistent with universality.


\section{Scaling in the unitary limit}

\label{sec_scal}

In this section we derive some scaling relations which relate observables at
different temperatures in the unitary limit. \ Our derivation is equivalent to
the treatment in \cite{Ho:2004b}. \ For the analysis it is simplest to first
take the temporal lattice spacing $a_{t}\rightarrow0$, and then take the
spatial lattice spacing $a\rightarrow0$. \ When we take the temporal lattice
spacing $a_{t}\rightarrow0$, we end up with a Hamiltonian lattice formulation,%
\begin{align}
H-\mu N  &  =\sum_{\vec{n}_{s},i}\left[  (-\mu+\tfrac{3}{m_{N}})a_{i}%
^{\dagger}(\vec{n}_{s})a_{i}(\vec{n}_{s})\right] \nonumber\\
&  -\tfrac{1}{2m_{N}}\sum_{\vec{n},i}\sum_{l_{s}=1,2,3}\left[  a_{i}^{\dagger
}(\vec{n}_{s})a_{i}(\vec{n}+\hat{l}_{s})+a_{i}^{\dagger}(\vec{n}_{s}%
)a_{i}(\vec{n}_{s}-\hat{l}_{s})\right] \nonumber\\
&  +C\sum_{\vec{n}_{s}}a_{\uparrow}^{\dagger}(\vec{n}_{s})a_{\uparrow}(\vec
{n}_{s})a_{\downarrow}^{\dagger}(\vec{n}_{s})a_{\downarrow}(\vec{n}_{s}).
\end{align}
Here, $a_{i}(\vec{n}_{s})$ is an annihilation operator for a neutron with spin
index $i$ at the spatial lattice site $\vec{n}_{s}$ and $C$ is the coupling constant.

In the unitary limit, which corresponds with $a\rightarrow0$ and
$a_{\text{scatt}}\rightarrow\infty$, it is not difficult to show that%
\begin{equation}
C=-\frac{\eta}{m_{N}}\text{,}%
\end{equation}
where $\eta$ is a constant,%
\begin{align}
\eta &  =\lim_{L\rightarrow\infty}\frac{L^{3}}{\sum_{\vec{k}\neq0\text{
}\operatorname{integer}}\frac{1}{\Omega_{\vec{k}}}}\simeq3.957,\\
\Omega_{_{\vec{k}}}  &  =6-2\cos\frac{2\pi k_{1}}{L}-2\cos\frac{2\pi k_{2}}%
{L}-2\cos\frac{2\pi k_{3}}{L}.
\end{align}
The derivation is as follows. \ At zero temperature, the value of $C$ can be
set by the condition that the two-particle scattering pole occurs at the
energy prescribed by L\"{u}scher's formula for energy levels in a periodic box
of length $L$ \cite{Luscher:1986pf,Beane:2003da}. \ In the actual lattice
simulations we describe later we will determine $C$ in a slightly different
way, but the two procedures agree in the limit that the lattice spacing goes
to zero. \ If we place the two-particle scattering pole at energy
$E_{\text{pole}}$ then the condition on $C$ is%
\begin{equation}
-\frac{1}{C}=\lim_{L\rightarrow\infty}\frac{1}{L^{3}}\sum_{\vec{k}\text{
}\operatorname{integer}}\frac{1}{-E_{\text{pole}}+\frac{1}{m_{N}}\Omega
_{\vec{k}}}. \label{pole}%
\end{equation}
L\"{u}scher's formula gives%
\begin{equation}
E_{\text{pole}}=\dfrac{4\pi a_{\text{scatt}}}{m_{N}L^{3}}[1+O\left(
\frac{a_{\text{scatt}}}{L}\right)  ]\text{,}%
\end{equation}
We keep $a_{\text{scatt}}$ finite for the moment. \ In the limit
$L\rightarrow\infty$ we split the sum in (\ref{pole})\ into the term $\vec
{k}=0$ and the remaining terms $\vec{k}\neq0$,%
\begin{align}
-\frac{1}{C}  &  =-\frac{1}{L^{3}}\frac{1}{E_{\text{pole}}}+\frac{1}{L^{3}%
}\sum_{\vec{k}\neq0\text{ }\operatorname{integer}}\frac{m_{N}}{\Omega
_{_{\vec{k}}}}\nonumber\\
&  =-\frac{m_{N}}{4\pi a_{\text{scatt}}}+\frac{m_{N}}{L^{3}}\sum_{\vec{k}%
\neq0\text{ }\operatorname{integer}}\frac{1}{\Omega_{_{\vec{k}}}}.
\end{align}
Therefore
\begin{equation}
C=\frac{1}{m_{N}}\frac{1}{\frac{1}{4\pi a_{\text{scatt}}}-\frac{1}{L^{3}}%
\sum_{\vec{k}\neq0\text{ }\operatorname{integer}}\frac{1}{\Omega_{_{\vec{k}}}%
}}.
\end{equation}
In the unitary limit where $a_{\text{scatt}}\rightarrow\infty$, we have%
\begin{equation}
C=-\frac{1}{m_{N}}\frac{L^{3}}{\sum_{\vec{k}\neq0\text{ }%
\operatorname{integer}}\frac{1}{\Omega_{_{\vec{k}}}}}=-\frac{\eta}{m_{N}},
\end{equation}
which corresponds with the three-dimensional attractive Hubbard model
\cite{Hubbard:1963,Sewer:2002,dosSantos:1994} with%
\begin{equation}
\frac{\left\vert U\right\vert }{t}=2\eta\simeq7.914.
\end{equation}

The grand canonical partition function for our system is
\begin{equation}
Z_{G}=Tr\exp\left[  -\beta(H-\mu N)\right]  ,
\end{equation}
or%
\begin{equation}
Z_{G}=Tr\exp\left[
\begin{array}
[c]{c}%
\sum_{\vec{n},i}\left[  (\beta\mu-\tfrac{3\beta}{m_{N}})a_{i}^{\dagger}%
(\vec{n})a_{i}(\vec{n})\right]  \\
+\tfrac{\beta}{2m_{N}}\sum_{\vec{n},i}\sum_{l=1,2,3}\left[  a_{i}^{\dagger
}(\vec{n})a_{i}(\vec{n}+\hat{l})+a_{i}^{\dagger}(\vec{n})a_{i}(\vec{n}-\hat
{l})\right]  \\
+\frac{\beta\eta}{m_{N}}\sum_{\vec{n}}a_{\uparrow}^{\dagger}(\vec
{n})a_{\uparrow}(\vec{n})a_{\downarrow}^{\dagger}(\vec{n})a_{\downarrow}%
(\vec{n}),
\end{array}
\right]
\end{equation}
We observe that $Z_{G}$ is a function of only two parameters, $\tfrac{\beta
}{2m_{N}}$ and $\beta\mu$. \ In terms of physical units, these parameters are%
\begin{equation}
\frac{\beta}{2m_{N}}=\frac{1}{2m_{N}^{\text{phys}}T^{\text{phys}}a^{2}}%
=\frac{1}{4\pi}\left(  \lambda_{T}^{\text{phys}}a^{-1}\right)  ^{2}\text{,}%
\end{equation}%
\begin{equation}
\beta\mu=\frac{\mu^{\text{phys}}}{T^{\text{phys}}}=\ln z,
\end{equation}
where $z$ is the fugacity,%
\begin{equation}
z=e^{\beta\mu}\text{.}%
\end{equation}
We can take the two independent parameters to be $\lambda_{T}^{\text{phys}%
}a^{-1}$ and $z$.

Consider any operator $F(a_{i},a_{j}^{\dagger})$ built from the lattice
annihilation and creation operators. \ We assume that%
\begin{equation}
\left\langle F\right\rangle _{\text{continuum}}\equiv\lim_{a\rightarrow
0}\left[  a^{D}\cdot\left\langle F(a_{i},a_{j}^{\dagger})\right\rangle
\right]  \label{power}%
\end{equation}
has a nonzero continuum limit for some power $D$. \ We know that
\begin{equation}
\left\langle F(a_{i},a_{j}^{\dagger})\right\rangle =\frac{Tr\left[
F(a_{i},a_{j}^{\dagger})\exp\left[  -\beta(H-\mu N)\right]  \right]
}{Tr\left[  \exp\left[  -\beta(H-\mu N)\right]  \right]  }%
\end{equation}
is a function of only $\lambda_{T}^{\text{phys}}a^{-1}$ and $z$. \ So in order
that the factor of $a^{D}$ drops out of (\ref{power}), we need in the
continuum limit%
\begin{equation}
\left\langle F(a_{i},a_{j}^{\dagger})\right\rangle \rightarrow\left(
\lambda_{T}^{\text{phys}}a^{-1}\right)  ^{D}f(z),
\end{equation}
where $f(z)$ is some function of the fugacity only. \ Hence%
\begin{equation}
\left\langle F\right\rangle _{\text{continuum}}=\left(  \lambda_{T}%
^{\text{phys}}\right)  ^{D}f(z)
\end{equation}
and so%
\begin{equation}
\left(  \lambda_{T}^{\text{phys}}\right)  ^{-D}\left\langle F\right\rangle
_{\text{continuum}}%
\end{equation}
is a function of $z$ only. \ We extend this to the case when the operator has
an explicit dependence on the displacement, $\vec{x}^{\text{phys}}.$ \ In that
case%
\begin{equation}
\left\langle F(\vec{x}^{\text{phys}})\right\rangle _{\text{continuum}}%
\equiv\lim_{a\rightarrow0}\left[  a^{D}\cdot\left\langle F(a_{i}%
,a_{j}^{\dagger},\vec{x}^{\text{phys}}a^{-1})\right\rangle \right]  ,
\end{equation}
where $\vec{x}^{\text{phys}}a^{-1}$ is the displacement in lattice units.
\ Therefore%
\begin{equation}
\left(  \lambda_{T}^{\text{phys}}\right)  ^{-D}\left\langle F(\vec
{x}^{\text{phys}})\right\rangle _{\text{continuum}}=f\left(  z,\vec
{x}^{\text{phys}}(\lambda_{T}^{\text{phys}})^{-1}\right)  .
\end{equation}

As an example we show that the particle density times three powers of the
thermal wavelength is a function of only the fugacity. \ The particle density
in the continuum limit is%
\begin{equation}
\left\langle \rho\right\rangle _{\text{continuum}}=\lim_{a\rightarrow0}\left[
a^{-3}\cdot\left\langle a_{\uparrow}^{\dagger}(\vec{n}_{s})a_{\uparrow}%
(\vec{n}_{s})+a_{\downarrow}^{\dagger}(\vec{n}_{s})a_{\downarrow}(\vec{n}%
_{s})\right\rangle \right]  ,
\end{equation}
where the expectation value can be measured at any spatial lattice site
$\vec{n}_{s}$. \ Therefore $D=-3$ and%
\begin{equation}
\left(  \lambda_{T}^{\text{phys}}\right)  ^{3}\left\langle \rho\right\rangle
_{\text{continuum}}%
\end{equation}
is a function of only the fugacity. \ Similarly the energy per particle times
inverse temperature,%
\begin{equation}
\beta\frac{E}{A}\text{;}%
\end{equation}
pressure times inverse temperature and three powers of the thermal
wavelength,
\begin{equation}
\left(  \lambda_{T}^{\text{phys}}\right)  ^{3}\beta P;
\end{equation}
and Fermi energy times inverse temperature,%
\begin{equation}
\beta E_{F}=\beta\frac{(3\pi^{2}\rho)^{2/3}}{2m_{N}};
\end{equation}
are all functions of fugacity only.

Using the unitary limit scaling relations we can derive a simple relation
between the energy density and pressure \cite{Ho:2004b}. \ The energy density
is
\begin{align}
\frac{E}{V}  &  =-\frac{1}{V}\frac{\partial}{\partial\beta}\ln Z_{G}+\mu
\rho\nonumber\\
&  =-\frac{1}{V}\left[  \frac{\partial}{\partial\beta}-\frac{\mu}{\beta}%
\frac{\partial}{\partial\mu}\right]  \ln Z_{G}.
\end{align}
We note that%
\begin{equation}
\left(  \lambda_{T}^{\text{phys}}\right)  ^{3}\frac{1}{V}\ln Z_{G}=\left(
\lambda_{T}^{\text{phys}}\right)  ^{3}\beta P,
\end{equation}
which is only a function of fugacity. \ \ Since%
\begin{equation}
\left[  \frac{\partial}{\partial\beta}-\frac{\mu}{\beta}\frac{\partial
}{\partial\mu}\right]  z=\left[  \frac{\partial}{\partial\beta}-\frac{\mu
}{\beta}\frac{\partial}{\partial\mu}\right]  e^{\beta\mu}=0,
\end{equation}
we have the result%
\begin{align}
\frac{E}{V}  &  =-\frac{1}{V}\left[  \frac{\partial}{\partial\beta}-\frac{\mu
}{\beta}\frac{\partial}{\partial\mu}\right]  \ln Z_{G}\nonumber\\
&  =-\frac{1}{V}\left(  \lambda_{T}^{\text{phys}}\right)  ^{3}\frac{1}{V}\ln
Z_{G}\left[  \frac{\partial}{\partial\beta}-\frac{\mu}{\beta}\frac{\partial
}{\partial\mu}\right]  \left[  \left(  \lambda_{T}^{\text{phys}}\right)
^{-3}\right] \nonumber\\
&  =\frac{3}{2\beta}\frac{1}{V}\ln Z_{G}=\frac{3}{2}P. \label{energydensity}%
\end{align}


\section{Results}

\label{sec_res}

In the following we present lattice simulation results for cold dilute neutron
matter in the unitary limit. \ We use a spatial lattice spacing of $a=(50$
MeV$)^{-1}$ and temporal lattice spacing of $a_{t}=(24$ MeV$)^{-1}$. \ The
temporal lattice spacing is sufficiently small that the results are close to
the $a_{t}\rightarrow0$ limit. \ We use the hybrid Monte Carlo algorithm
\cite{Duane:1987de} to generate Hubbard-Stratonovich field configurations as
described in \cite{Lee:2004qd}. \ We use diagonal preconditioning before each
conjugate gradient solve as described in [I].

The finite volume error was tested by going to larger volumes, and the final
lattice sizes were chosen so that the finite volume error was less than one
percent. \ The lattice dimensions we used for the various temperatures are
shown in Table 1.
\[%
\genfrac{}{}{0pt}{0}{\text{Table 1: \ Lattice volumes for various
temperatures}}{%
\begin{tabular}
[c]{|l|l|l|}\hline
$T\text{ (MeV})$ & $L$ & $L_{t}$\\\hline
$4$ & $4$ & $6$\\\hline
$3$ & $5$ & $8$\\\hline
$2$ & $5$ & $12$\\\hline
$1.5$ & $6$ & $16$\\\hline
$1$ & $6$ & $24$\\\hline
\end{tabular}
}%
\]

We adjust the coefficient $C$ of the four-fermion interaction as suggested in
[I]. \ For each temperature and lattice volume, we compute the density for
both free lattice fermions and for lattice regularized two-particle bubbles.
\ We use the free fermion result in order to determine the first virial
coefficient $b_{1}(T)$ (see Table 1 in [I]) and the bubble sum to compute the
second virial coefficient $b_{2}(T),$%
\begin{equation}
\rho^{\text{bubble}}\approx\frac{2}{\lambda_{T}^{3}}b_{1}(T)\left[  z+2\cdot
b_{2}(T)z^{2}\right]  \text{.}%
\end{equation}
We then adjust $C$ so that
\begin{equation}
b_{2}(T)=3\cdot2^{-\frac{5}{2}}\text{.}%
\end{equation}
This constraint is also used to compute the derivative of $C$ with respect to
the temporal lattice spacing. \ The results for $C(T)$ and $\frac{dC}%
{d\alpha_{t}}$ are shown in Table 2.
\[%
\genfrac{}{}{0pt}{0}{\text{Table 2: \ }C\text{ and }\frac{dC}{d\alpha_{t}%
}\text{ on the lattice}}{%
\begin{tabular}
[c]{|l|l|l|}\hline
$T$ (MeV) & $C$ ($10^{-4}$ MeV$^{-2}$) & $\frac{dC}{d\alpha_{t}}$ ($10^{-5}$
MeV$^{-2}$)\\\hline
$4$ & $-0.971$ & $-0.76$\\\hline
$3$ & $-0.948$ & $-1.40$\\\hline
$2$ & $-0.958$ & $-2.12$\\\hline
$1.5$ & $-0.987$ & $-2.47$\\\hline
$1$ & $-1.043$ & $-2.68$\\\hline
$0.667$ & $-1.098$ & $-2.65$\\\hline
$0.5$ & $-1.128$ & $-2.52$\\\hline
\end{tabular}
}%
\]
For each temperature we have probed densities up to a quarter-filled lattice.
\ With a spatial lattice spacing of $(50$ MeV$)^{-1}$, the quarter-filled
lattice corresponds with a density of $0.0081$ fm$^{-3}$. \ Beyond this one
might find significant lattice artifacts. \ 


\subsection{Density versus fugacity}

\label{sec_dens}

\begin{figure}
[ptb]
\begin{center}
\includegraphics[
height=3.966in,
width=2.7864in,
angle=-90
]%
{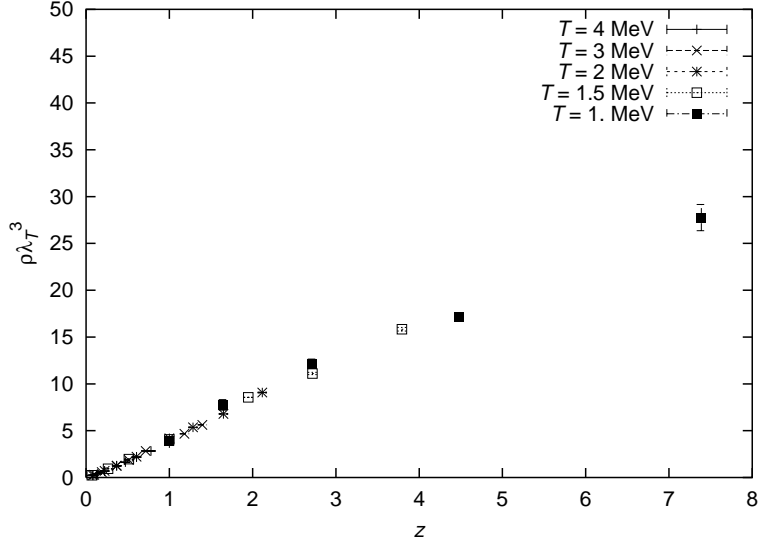}%
\caption{Density times $\lambda_{T}^{3}$ versus fugacity for various
temperatures.}%
\label{rho_z}%
\end{center}
\end{figure}

\begin{figure}
[ptbptb]
\begin{center}
\includegraphics[
height=3.966in,
width=2.7864in,
angle=-90
]%
{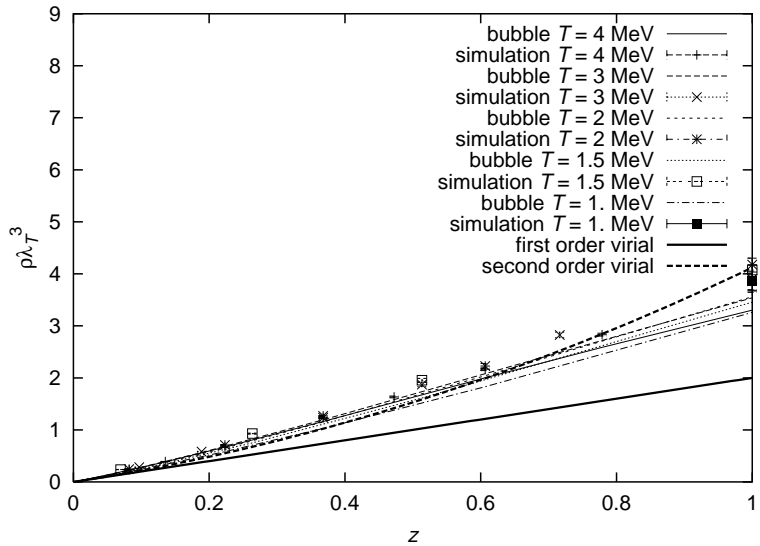}%
\caption{Density times $\lambda_{T}^{3}$ versus fugacity for small fugacity.
\ We also include comparisons with the virial expansion at first and second
orders.}%
\label{rho_z_zoom}%
\end{center}
\end{figure}

In Fig. \ref{rho_z} we plot the density times $\lambda_{T}^{3}$ versus
fugacity for temperatures $T=4$, $3,$ $2,$ $1.5,$ and $1$ MeV.
%
The data from the five different temperatures appear to fall on a single
curve. \ This suggests that $\rho\lambda_{T}^{3}$ depends only on $z,$ as
predicted by unitary limit scaling. \ In Fig. \ref{rho_z_zoom} we magnify the
plot of $\rho\lambda_{T}^{3}$ at low fugacity, showing both bubble chain
results and full simulation results.
%
We show the first order and second order virial results with%
\begin{equation}
b_{2}(T)=3\cdot2^{-\frac{5}{2}}\approx0.530.
\end{equation}
Since we have tuned the interaction coefficient to produce the correct second
order virial coefficient, it is not surprising that the lattice data agrees
with the virial expansion at low fugacity.


\subsection{Energy per particle versus fugacity}

\label{E/A}

\begin{figure}
[ptb]
\begin{center}
\includegraphics[
height=3.966in,
width=2.7864in,
angle=-90
]%
{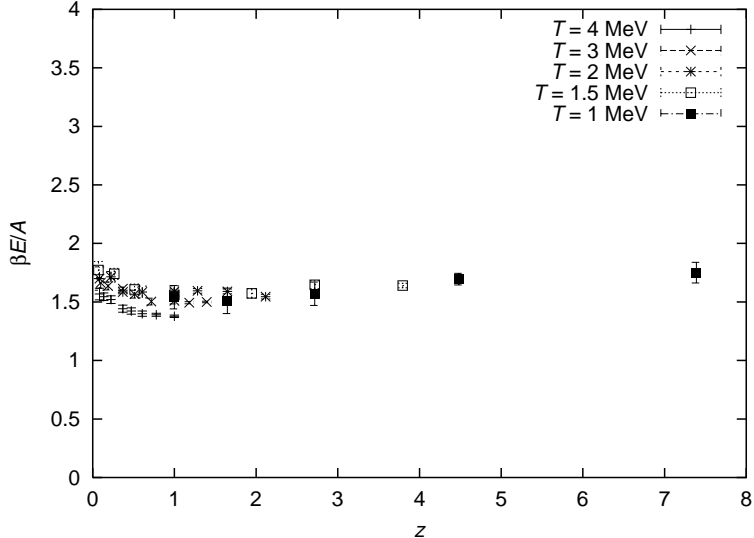}%
\caption{Energy per particle times $\beta$ versus fugacity. \ We show results
for $T=4$, $3,$ $2,$ $1.5,$ and $1$ MeV.}%
\label{energy_z}%
\end{center}
\end{figure}

In Fig. \ref{energy_z} we plot the energy per particle times $\beta$ versus
fugacity for temperatures $T=4$, $3,$ $2,$ $1.5,$ and $1$ MeV.
%
The energy per particle times $\beta$ appears to depends only on fugacity, as
predicted by scaling in the unitary limit. \ The small deviation for different
temperatures appears to be due mainly to an overall shift in the height of the
curves. \ In the continuum limit at $z=0$, we expect the equipartition result%
\begin{equation}
\frac{\beta E}{A}=\frac{3}{2}.
\end{equation}
The slight deviations from $\frac{3}{2}$ at $z=0$ can be attributed to lattice
cutoff effects in the free particle kinetic energy.

\begin{figure}
[ptb]
\begin{center}
\includegraphics[
height=3.966in,
width=2.7864in,
angle=-90
]%
{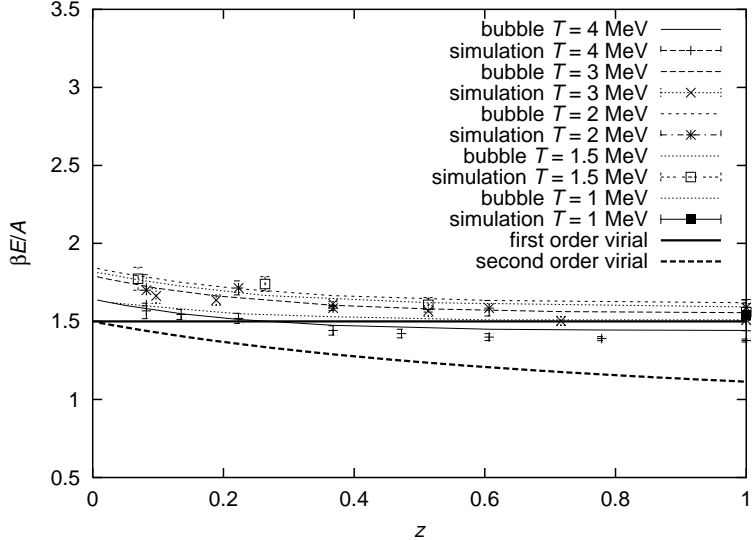}%
\caption{Energy per particle times $\beta$ versus fugacity at small fugacity.
\ We compare with first and second order virial expansion results.}%
\label{energy_z_zoom}%
\end{center}
\end{figure}

In Fig. \ref{energy_z_zoom} we magnify the same plot for small fugacity and
include both bubble chain calculation results and full simulation results.
%
We also show the first and second order virial results in the continuum. \ At
first order we have%
\begin{equation}
\frac{\beta E}{A}=\frac{3}{2},
\end{equation}
and at second order,%
\begin{equation}
\frac{\beta E}{A}=\frac{\left(  \frac{3}{2}-\ln z\right)  z+\frac{3}{4\sqrt
{2}}\left(  \frac{3}{2}-2\ln z\right)  z^{2}}{z+\frac{3}{2\sqrt{2}}z^{2}}+\ln
z.
\end{equation}
We see that apart from small shifts in the overall height, the lattice results
agree with the continuum virial results.


\subsection{Energy density and pressure}

\label{EvsP}

\begin{figure}
[ptb]
\begin{center}
\includegraphics[
height=3.966in,
width=2.7864in,
angle=-90
]%
{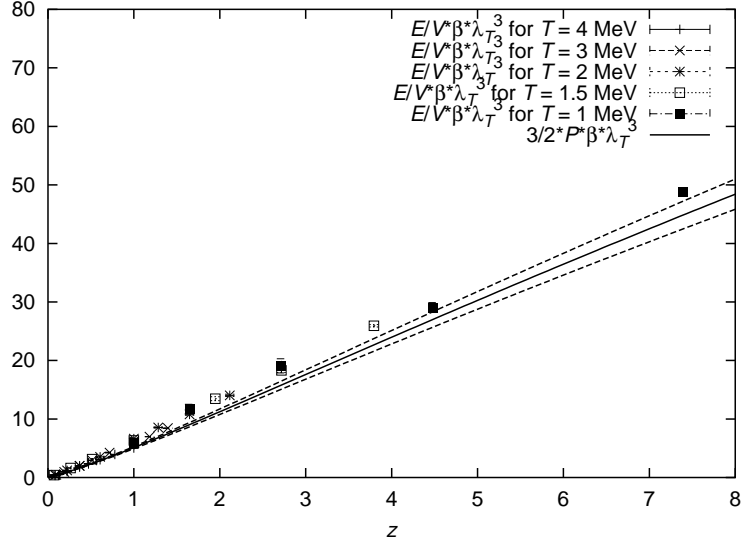}%
\caption{Energy density times $\beta\lambda_{T}^{3}$ versus fugacity for
various temperatures. \ We compare with the pressure times $\frac{3}{2}%
\beta\lambda_{T}^{3}.$}%
\label{scaled_eoverv_p}%
\end{center}
\end{figure}

In Fig. \ref{scaled_eoverv_p} we show the energy density times $\beta
\lambda_{T}^{3}$ versus fugacity for temperatures $T=4$, $3,$ $2,$ $1.5,$ and
$1$ MeV. \ Scaling in the unitary limit requires that the energy density times
$\beta\lambda_{T}^{3}$ is only a function of fugacity, and this appears to be
the case.
%
In Fig. \ref{scaled_eoverv_p} we also plot the pressure times $\frac{3}%
{2}\beta\lambda_{T}^{3}$, which according to (\ref{energydensity}) should
equal the energy density times $\beta\lambda_{T}^{3}$. \ We have computed the
pressure by numerical integration of the density as a function of chemical
potential,%
\begin{equation}
P=\frac{T}{V}\ln Z_{G}=\frac{1}{V}\int_{-\infty}^{\mu}A(\mu^{\prime}%
)d\mu^{\prime}=\int_{-\infty}^{\mu}\rho(\mu^{\prime})d\mu^{\prime}.
\end{equation}
We see that the lattice results appear to confirm the unitary limit relation,%
\begin{equation}
\frac{E}{V}=\frac{3}{2}P.
\end{equation}


\subsection{Reduced energy versus reduced temperature}

\label{sec_evst}

\begin{figure}
[ptb]
\begin{center}
\includegraphics[
height=3.966in,
width=2.7864in,
angle=-90
]%
{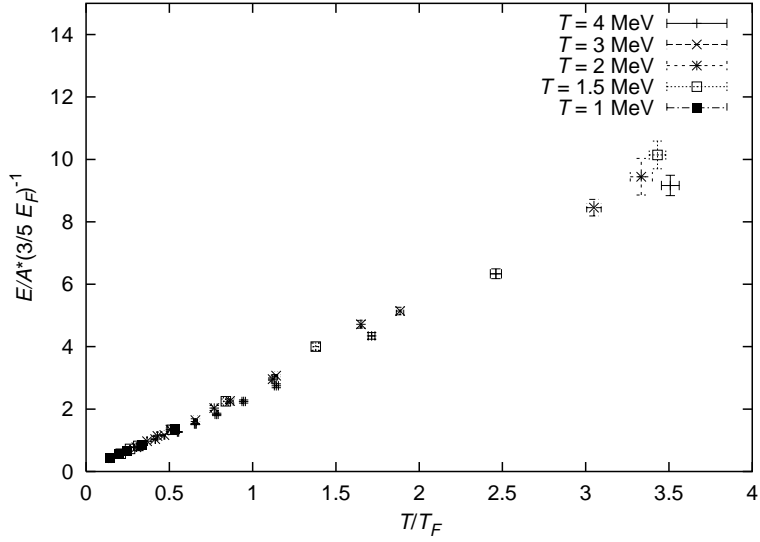}%
\caption{The energy per particle divided by $\frac{3}{5}E_{F}$ versus
temperature divided by $T_{F}$ for temperatures $T=4$, $3,$ $2,$ $1.5,$ and
$1$ MeV.}%
\label{e_t}%
\end{center}
\end{figure}

\begin{figure}
[ptbptb]
\begin{center}
\includegraphics[
height=3.966in,
width=2.7864in,
angle=-90
]%
{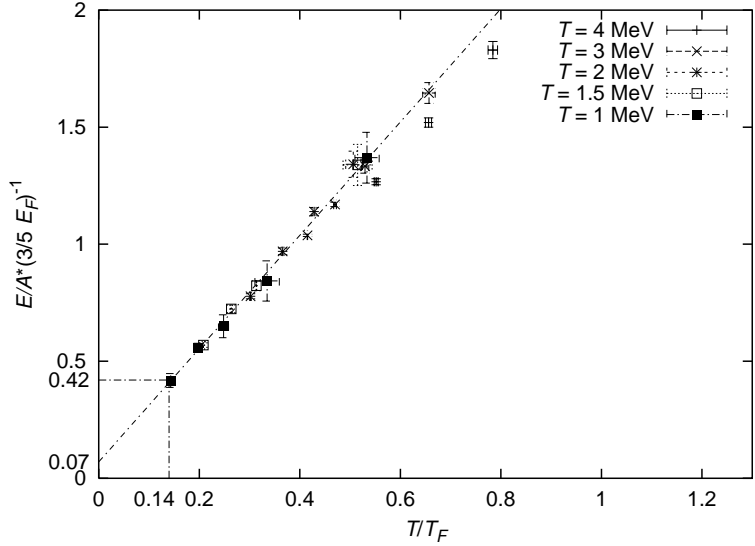}%
\caption{A magnified plot at low temperatures of the energy per particle
divided by $\frac{3}{5}E_{F}$ versus temperature divided by $T_{F}$.}%
\label{e_t_zoom}%
\end{center}
\end{figure}

In units where Boltzmann's constant equals $1,$ the degeneracy temperature
$T_{F}$ is the same as the Fermi energy,%
\begin{equation}
T_{F}=E_{F}=\frac{(3\pi^{2}\rho)^{2/3}}{2m_{N}}.
\end{equation}
In Fig. \ref{e_t} we plot the energy per particle divided by $\frac{3}{5}%
E_{F}$ versus the temperature divided by $T_{F}$ for temperatures $T=4$, $3,$
$2,$ $1.5,$ and $1$ MeV. \ As expected from unitary limit scaling all points
appear to lie on a single curve.
%
In Fig. \ref{e_t_zoom} we show a magnified plot of the energy per particle
divided by $\frac{3}{5}E_{F}$ versus the temperature divided by $T_{F}$. \ The
data points at the lowest values of $T/T_{F}$ appear to lie on straight line
with intercept
\begin{equation}
\frac{E/A}{\frac{3}{5}E_{F}}=0.07.
\end{equation}
One expects the curve to reach $T/T_{F}=0$ with zero slope. \ The actual
intercept at $T/T_{F}=0,$ which corresponds with the parameter $\xi$ in the
zero temperature relation,%
\begin{equation}
\frac{E}{A}=\xi\frac{3}{5}\frac{k_{F}^{2}}{2m}\text{,} \label{Bertsch}%
\end{equation}
should likely be somewhere between $0.07$ and $0.42$.
%
Since there is no noticeable nonanalytic behavior in the data shown, this
suggests that the superfluid critical temperature $T_{C}$ should satisfy%
\begin{equation}
\frac{T_{C}}{T_{F}}<0.14.
\end{equation}


\subsection{Axial response and spin susceptibility}

\label{sec_sus}

The axial response at zero momentum is defined as \cite{Burrows:1998cg}%
\begin{equation}
S_{\text{A}}(0)=\frac{1}{\left\langle N\right\rangle }\left\langle
(N_{\uparrow}-N_{\downarrow})(N_{\uparrow}-N_{\downarrow})\right\rangle ,
\end{equation}
where $N_{\uparrow(\downarrow)}$ is the number operator for spin up(down)
neutrons. $\ N$ is the total number operator so that%
\begin{equation}
\left\langle N\right\rangle =\left\langle N_{\uparrow}+N_{\downarrow
}\right\rangle =A.
\end{equation}
$S_{\text{A}}(0)$ is a dimensionless quantity with a finite continuum limit
and therefore should be a function of fugacity only. \ It is normalized to
equal $1$ at zero density. $\ S_{\text{A}}(0)$ is proportional to the Pauli
spin susceptibility \cite{dosSantos:1994,Sewer:2002},%
\begin{equation}
\chi_{P}=\frac{1}{\left\langle N\right\rangle T}\sum_{\vec{n}_{s},\vec{n}%
_{s}^{\prime}}\left\langle \left[  a_{i}^{\dagger}(\vec{n}_{s})[\vec{\sigma
}]_{ij}a_{j}(\vec{n}_{s})\right]  \cdot\left[  a_{k}^{\dagger}(\vec{n}%
_{s}^{\prime})[\vec{\sigma}]_{kl}a_{l}(\vec{n}_{s}^{\prime})\right]
\right\rangle =\frac{3S_{\text{A}}(0)}{T}.
\end{equation}

\begin{figure}
[ptb]
\begin{center}
\includegraphics[
height=3.966in,
width=2.7864in,
angle=-90
]%
{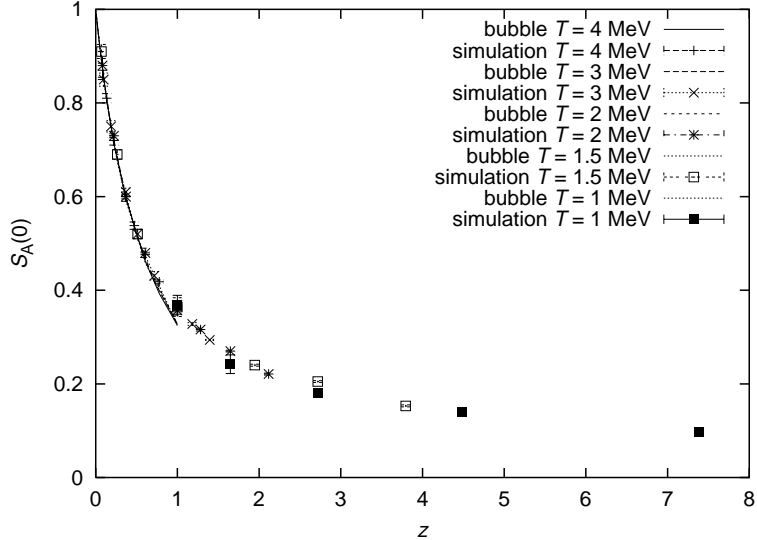}%
\caption{Lattice results for the axial response as a function of fugacity for
temperatures $T=4$, $3,$ $2,$ $1.5,$ and $1$ MeV.}%
\label{a0_z}%
\end{center}
\end{figure}

In Fig. \ref{a0_z} we show the axial response as a function of fugacity for
temperatures $T=4$, $3,$ $2,$ $1.5,$ and $1$ MeV.%
%
The lattice results for the axial response appear to lie on one curve, as
predicted by the unitary limit scaling relations.

The decrease in the axial response indicates that the transfer of spin from
one spatial region to another is being suppressed. \ This is due to the
formation of spin zero pairs. \ A superfluid transition leads to nonanalytic
behavior in the susceptibility. \ A sharp crossover in the susceptibility can
be used to define a pseudogap phase. \ We do not observe any of these
phenomena. \ The susceptibility becomes quite small for the most degenerate
systems studied, but the dependence on temperature is smooth.


\subsection{Dineutron correlation function}

\label{sec_cor}

\begin{figure}
[ptb]
\begin{center}
\includegraphics[
height=3.966in,
width=2.7864in,
angle=-90
]%
{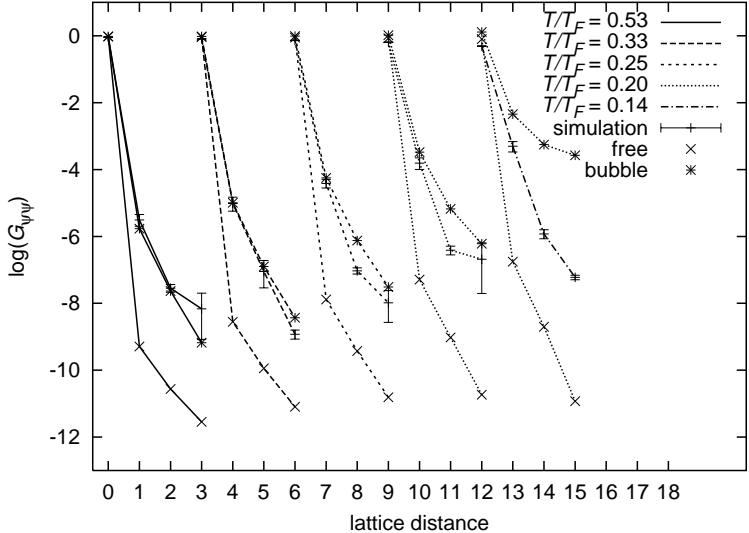}%
\caption{Logarithm of the dineutron correlation function versus lattice
distance measured along a lattice axis. \ We show results for $T=1$\ MeV$.$}%
\label{dineutron}%
\end{center}
\end{figure}

We define the equal-time dineutron correlation function as%
\begin{equation}
G_{\psi\psi}(\vec{n}_{s})=\left\langle a_{\downarrow}(\vec{n}_{s})a_{\uparrow
}(\vec{n}_{s})a_{\uparrow}^{\dagger}(0)a_{\downarrow}^{\dagger}%
(0)\right\rangle .
\end{equation}
In Fig. \ref{dineutron} we show the logarithm of the dineutron correlation
function as a function of lattice distance measured along a lattice axis. \ We
show results for $T=1$\ MeV and various values of $T/T_{F}$. \ We have
staggered the plots for better viewing, and the five highest points represent
data measured at zero lattice distance.
%
For comparison we also show the free dineutron correlation function as well as
the result of the bubble chain approximation. \ We observe that the
interacting correlation function is larger than the non-interacting one for
all temperatures. \ This reflects the attractive interaction in the spin zero
channel. \ At the higher temperatures the correlation function is
quantitatively explained by the bubble chain approximation.\ At $T/T_{F}=0.14$
the bubble chain correlator starts to level off, indicative of long range
order, but the full correlation function does not. \ This implies that
$T_{C}/T_{F}<0.14$.


\section{Summary}

\label{sec_sum}

We have studied neutron matter in the unitary limit on the lattice. \ In [I]
we concentrated on the low density/high temperature equation of state and
compared the results to the virial expansion. \ We found significant finite
lattice spacing errors and suggested that the scaling behavior can be improved
by tuning the second virial coefficient rather than the zero temperature
scattering length.

In part II we showed that tuning the second virial coefficient improves
universal scaling for the entire range of densities and temperatures studied.
\ We found, in particular, that the energy per particle in units of $T_{F}$
only depends on $\mu/T$ and that the energy density is 3/2 times the pressure.
\ In the temperature range studied the energy per particle in units of $T_{F}$
is essentially linear in $T/T_{F}$. \ This means that at present we can only
provide bounds on the universal parameter $\xi$. \ We find $0.07\leq\xi
\leq0.42$. \ This result is marginally consistent with the Green function
Monte Carlo result $\xi=0.44$ \cite{Carlson:2003z}.

We also studied the spin susceptibility and the dineutron correlation
function. \ We find that the spin susceptibility is strongly suppressed in the
most degenerate systems, but no clear phase transition is observed. \ We also
find that the dineutron correlation function is strongly enhanced over the
free correlation function, but no long range order is observed. \ We conclude
that the critical temperature for the the transition to a superfluid phase is
less than $0.14T_{F}$. \ This bound is lower than the result
$T_{C}=0.22(3)T_{F}$ \cite{Bulgac:2005a} but consistent with the result
$T_{C}=0.035(4)T_{F}$ obtained in \cite{Wingate:2005xy}. \ Green's function
Monte Carlo calculations are restricted to $T=0$ and can only determine the
gap, not the critical temperature. \ Carlson, et. al.,~\cite{Carlson:2003z}
find $\Delta=0.9\cdot\frac{3}{5}E_{F}$. \ If the critical temperature were
related to the gap by $T_{C}=0.57\Delta$ as in BCS theory this would imply
$T_{C}=0.31T_{F}$, but there is no reason to expect this relation to hold in
the unitary limit.

Acknowledgments: This work is supported in part by the US Department of Energy
grants DE-FG-88ER40388 (T.S.) and DE-FG02-04ER41335 (D.L.).

\bibliographystyle{apsrev}
\bibliography{NuclearMatter}

\begin{thebibliography}{12}
\expandafter\ifx\csname natexlab\endcsname\relax\def\natexlab#1{#1}\fi
\expandafter\ifx\csname bibnamefont\endcsname\relax
  \def\bibnamefont#1{#1}\fi
\expandafter\ifx\csname bibfnamefont\endcsname\relax
  \def\bibfnamefont#1{#1}\fi
\expandafter\ifx\csname citenamefont\endcsname\relax
  \def\citenamefont#1{#1}\fi
\expandafter\ifx\csname url\endcsname\relax
  \def\url#1{\texttt{#1}}\fi
\expandafter\ifx\csname urlprefix\endcsname\relax\def\urlprefix{URL }\fi
\providecommand{\bibinfo}[2]{#2}
\providecommand{\eprint}[2][]{\url{#2}}

\bibitem[{\citenamefont{Ho}(2004)}]{Ho:2004b}
\bibinfo{author}{\bibfnamefont{T.-L.} \bibnamefont{Ho}},
  \bibinfo{journal}{Phys. Rev. Lett.} \textbf{\bibinfo{volume}{92}},
  \bibinfo{pages}{090402} (\bibinfo{year}{2004}).

\bibitem[{\citenamefont{L{\"u}scher}(1986)}]{Luscher:1986pf}
\bibinfo{author}{\bibfnamefont{M.}~\bibnamefont{L{\"u}scher}},
  \bibinfo{journal}{Commun. Math. Phys.} \textbf{\bibinfo{volume}{105}},
  \bibinfo{pages}{153} (\bibinfo{year}{1986}).

\bibitem[{\citenamefont{Beane et~al.}(2004)\citenamefont{Beane, Bedaque,
  Parreno, and Savage}}]{Beane:2003da}
\bibinfo{author}{\bibfnamefont{S.~R.} \bibnamefont{Beane}},
  \bibinfo{author}{\bibfnamefont{P.~F.} \bibnamefont{Bedaque}},
  \bibinfo{author}{\bibfnamefont{A.}~\bibnamefont{Parreno}}, \bibnamefont{and}
  \bibinfo{author}{\bibfnamefont{M.~J.} \bibnamefont{Savage}},
  \bibinfo{journal}{Phys. Lett.} \textbf{\bibinfo{volume}{B585}},
  \bibinfo{pages}{106} (\bibinfo{year}{2004}), \eprint{hep-lat/0312004}.

\bibitem[{\citenamefont{Hubbard}(1963)}]{Hubbard:1963}
\bibinfo{author}{\bibfnamefont{J.}~\bibnamefont{Hubbard}},
  \bibinfo{journal}{Proc. Roy. Soc.} \textbf{\bibinfo{volume}{276}},
  \bibinfo{pages}{238} (\bibinfo{year}{1963}).

\bibitem[{\citenamefont{Sewer et~al.}(2002)\citenamefont{Sewer, Zotos, and
  Beck}}]{Sewer:2002}
\bibinfo{author}{\bibfnamefont{A.}~\bibnamefont{Sewer}},
  \bibinfo{author}{\bibfnamefont{X.}~\bibnamefont{Zotos}}, \bibnamefont{and}
  \bibinfo{author}{\bibfnamefont{H.}~\bibnamefont{Beck}},
  \bibinfo{journal}{Phys. Rev. B} \textbf{\bibinfo{volume}{66}},
  \bibinfo{pages}{140504(R)} (\bibinfo{year}{2002}), \eprint{cond-mat/0204053}.

\bibitem[{\citenamefont{dos Santos}(1994)}]{dosSantos:1994}
\bibinfo{author}{\bibfnamefont{R.}~\bibnamefont{dos Santos}},
  \bibinfo{journal}{Phys. Rev. B} \textbf{\bibinfo{volume}{50}},
  \bibinfo{pages}{635} (\bibinfo{year}{1994}).

\bibitem[{\citenamefont{Duane et~al.}(1987)\citenamefont{Duane, Kennedy,
  Pendleton, and Roweth}}]{Duane:1987de}
\bibinfo{author}{\bibfnamefont{S.}~\bibnamefont{Duane}},
  \bibinfo{author}{\bibfnamefont{A.~D.} \bibnamefont{Kennedy}},
  \bibinfo{author}{\bibfnamefont{B.~J.} \bibnamefont{Pendleton}},
  \bibnamefont{and} \bibinfo{author}{\bibfnamefont{D.}~\bibnamefont{Roweth}},
  \bibinfo{journal}{Phys. Lett.} \textbf{\bibinfo{volume}{B195}},
  \bibinfo{pages}{216} (\bibinfo{year}{1987}).

\bibitem[{\citenamefont{Lee and Schaefer}(2005)}]{Lee:2004qd}
\bibinfo{author}{\bibfnamefont{D.}~\bibnamefont{Lee}} \bibnamefont{and}
  \bibinfo{author}{\bibfnamefont{T.}~\bibnamefont{Schaefer}},
  \bibinfo{journal}{Phys. Rev.} \textbf{\bibinfo{volume}{C72}},
  \bibinfo{pages}{024006} (\bibinfo{year}{2005}), \eprint{nucl-th/0412002}.

\bibitem[{\citenamefont{Burrows and Sawyer}(1998)}]{Burrows:1998cg}
\bibinfo{author}{\bibfnamefont{A.}~\bibnamefont{Burrows}} \bibnamefont{and}
  \bibinfo{author}{\bibfnamefont{R.~F.} \bibnamefont{Sawyer}},
  \bibinfo{journal}{Phys. Rev.} \textbf{\bibinfo{volume}{C58}},
  \bibinfo{pages}{554} (\bibinfo{year}{1998}), \eprint{astro-ph/9801082}.

\bibitem[{\citenamefont{Carlson et~al.}(2003)\citenamefont{Carlson, Chang,
  Pandharipande, and Schmidt}}]{Carlson:2003z}
\bibinfo{author}{\bibfnamefont{J.}~\bibnamefont{Carlson}},
  \bibinfo{author}{\bibfnamefont{S.~Y.} \bibnamefont{Chang}},
  \bibinfo{author}{\bibfnamefont{V.~R.} \bibnamefont{Pandharipande}},
  \bibnamefont{and} \bibinfo{author}{\bibfnamefont{K.}~\bibnamefont{Schmidt}},
  \bibinfo{journal}{Phys. Rev. Lett.} \textbf{\bibinfo{volume}{91}},
  \bibinfo{pages}{50401} (\bibinfo{year}{2003}), \eprint{physics/0303094}.

\bibitem[{\citenamefont{Bulgac et~al.}(2005)\citenamefont{Bulgac, Drut, and
  Magierski}}]{Bulgac:2005a}
\bibinfo{author}{\bibfnamefont{A.}~\bibnamefont{Bulgac}},
  \bibinfo{author}{\bibfnamefont{J.~E.} \bibnamefont{Drut}}, \bibnamefont{and}
  \bibinfo{author}{\bibfnamefont{P.}~\bibnamefont{Magierski}}
  (\bibinfo{year}{2005}), \eprint{cond-mat/0505374}.

\bibitem[{\citenamefont{Wingate}(2005)}]{Wingate:2005xy}
\bibinfo{author}{\bibfnamefont{M.}~\bibnamefont{Wingate}}
  (\bibinfo{year}{2005}), \eprint{cond-mat/0502372}.

\end{thebibliography}

\end{document}